\documentclass[12pt]{article}
\usepackage{wallpaper}
\usepackage[margin=1in, paperwidth=8.5in, paperheight=11in]{geometry}
\usepackage{amsmath}
\usepackage{graphicx}%
\usepackage{amsfonts}%
\usepackage{amssymb}
\usepackage{color}
\usepackage{float,subfigure}
\usepackage[round]{natbib}
\usepackage{cite}
\usepackage{setspace}
\usepackage{comment}
\usepackage{enumerate}
\usepackage{verbatim}
\usepackage{rotating}
\usepackage{microtype}
\usepackage[normalem]{ulem}
\usepackage{url}
\usepackage[hidelinks]{hyperref} 
\usepackage{cancel}
\usepackage{animate}
\usepackage{mathtools}
\usepackage{epsfig}
\usepackage{epstopdf}
\usepackage{multirow}

\usepackage{amsthm}

\hypersetup{pdfpagemode=UseNone}
\theoremstyle{plain}
\newtheorem{thm}{Theorem} 
\newtheorem*{thm*}{Theorem} 

\theoremstyle{definition}

\newcommand{\balpha}{ \mbox{\boldmath $ \alpha $} }
\newcommand{\bphi}{ \mbox{\boldmath $\phi$}}

\newcommand{\eps}{ \mbox{$\epsilon$}}

\newcommand{\btheta}{ \mbox{\boldmath $ \theta $} }

\newcommand{\blambda}{ \mbox{\boldmath $\lambda$} }

\newcommand{\bpsi}{ \mbox{\boldmath $\psi$}}

\newcommand{\ba}{ {\bf a} }

\newcommand{\bb}{ {\bf b} }

\newcommand{\bn}{ {\bf n} }

\newcommand{\bx}{ {\bf x} }

\newcommand{\by}{ {\bf y} }

\newcommand{\bz}{ {\bf z} }

\newcommand{\given}{\,\vert\,}

\newcommand{\Pois}{\mbox{$\text{Pois}$}}

\newcommand{\NegBin}{\mbox{$\text{NegBin}$}}

\newcommand{\Gam}{\mbox{$\text{Gamma}$}}

\newcommand{\Dir}{\mbox{Dir}}
\newcommand{\Mult}{\mbox{Mult}}

\newcommand{\bpi}{ \mbox{\boldmath $ \pi $} }

\begin{document}


\singlespacing
\thispagestyle{empty}
\setcounter{page}{0}

\begin{center}

{\Large \textbf{Improving the Utility of Poisson-Distributed, Differentially Private Synthetic Data via Prior Predictive Truncation with an Application to CDC~WONDER}} 

\bigskip

\textbf{Harrison Quick$^{*}$}\\ 
Department of Epidemiology and Biostatistics\\Drexel University, Philadelphia, PA 19104, USA\\
$^{*}$ \emph{email:} hsq23@drexel.edu

\end{center}

\noindent\textsc{Summary.}
CDC WONDER is a web-based tool for the dissemination of epidemiologic data collected by the National Vital Statistics System.  While CDC WONDER has built-in privacy protections, they do not satisfy formal privacy protections such as differential privacy and thus are susceptible to targeted attacks.  Given the importance of making high-quality public health data publicly available while preserving the privacy of the underlying data subjects, we aim to improve the utility of a recently developed approach for generating Poisson-distributed, differentially private synthetic data by using publicly available information to truncate the range of the synthetic data.  Specifically, we utilize county-level population information from the U.S.\ Census Bureau and national death reports produced by the CDC to inform prior distributions on county-level death rates and infer reasonable ranges for Poisson-distributed, county-level death counts.  In doing so, the requirements for satisfying differential privacy for a given privacy budget can be reduced by several orders of magnitude, thereby leading to substantial improvements in utility.  To illustrate our proposed approach, we consider a dataset comprised of over 26,000 cancer-related deaths from the Commonwealth of Pennsylvania belonging to over 47,000 combinations of cause-of-death and demographic variables such as age, race, sex, and county-of-residence and demonstrate the proposed framework's ability to preserve features such as geographic, urban/rural, and racial disparities present in the true data.

%
\vspace{12pt}

\newpage
\doublespacing
\section{Introduction}
CDC WONDER --- the CDC's Wide-ranging Online Data for Epidemiologic Research system --- is a web-based tool for the dissemination of epidemiologic data collected by the National Vital Statistics System.  Via CDC WONDER, researchers can gain immediate access to vital statistics data, such as the number of births and deaths stratified by geographic region (e.g., state, county), demographic variables (e.g., age, race, sex), specific causes of death (e.g., ICD codes), and by year from 1968 to 2019, subject to the suppression of small counts \citep{cdc:sharing}.  Unfortunately, suppression techniques like those implemented on CDC WONDER have been shown to be susceptible to targeted attacks \citep[e.g., ][]{dinur:nissim,holan2010,quick:zero}, motivating the development of alternative approaches to safely disseminate the nation's vital statistics data for public-use.

To this end, recent work has aimed at replacing the existing CDC WONDER with a ``Synthetic CDC WONDER'' in which all county-level counts would be replaced by \emph{synthetic} values generated in a Bayesian posterior predictive framework.  \citet{quick:synthetic} proposed the use of models from the disease mapping literature --- specifically, the conditional autoregressive (CAR) model of \citet{bym} and the multivariate CAR model of \citet{gelfand:mcar} --- to generate synthetic counts of the number of heart disease-related deaths in U.S.\ counties over a 10-year period across multiple age groups.  While this approach may produce synthetic data with high utility by virtue of estimating spatial-, temporal-, and between-age sources of dependence in the true data and preserving those dependencies in the synthetic data, the approach of \citet{quick:synthetic} has not been shown to satisfy formal privacy protections such as the definition of \emph{differential privacy} \citep{dwork:etal}.

In contrast, \citet{quick:diffpriv} set out to create a differentially private framework for generating synthetic data in the context of CDC WONDER.  Inspired by the work of \citet{onthemap} --- the methodological framework behind the U.S.\ Census Bureau's OnTheMap tool --- \citet{quick:diffpriv} established criteria in which synthetic data generated from the posterior predictive distribution of a Poisson-gamma model could satisfy $\eps$-differential privacy.  Specifically, the approach assumes the number of events in a given region arises from a Poisson distribution \citep[as is common in the disease mapping literature; e.g.,][]{brillinger} and incorporates external information regarding the size of the at-risk population and an estimate of the event rate in an effort to produce synthetic data with greater utility.  Furthermore, \citet{quick:diffpriv} demonstrates how the approach of \citet{onthemap} can be viewed as a special case of the proposed Poisson-gamma framework in which the underlying population sizes and event rates are assumed to be equal for all groups.

The drawback of the approach of \citet{quick:diffpriv} --- and the multinomial-Dirichlet approach of \citet{onthemap} that inspired it --- is that the worst case scenario underlying the criteria for satisfying $\eps$-differential privacy is highly unrealistic.  In particular, it limits the disclosure risk in the scenario in which a region that experienced only a single event in the true dataset is assigned \emph{all} of the events in the synthetic data.  In the context of data from CDC WONDER, this could imply that the number of deaths allocated to a region could far exceed the size of its at-risk population.  In this paper, we propose the use of \emph{prior predictive truncation} --- i.e., restricting the domain of the synthetic data based on the prior predictive distribution --- to improve the utility of our synthetic data.
Section~\ref{sec:methods} provides a brief overview of how synthetic data can be generated in a posterior predictive fashion and the approach of \citet{quick:diffpriv}, followed by a detailed description of our proposed prior predictive truncation approach. To illustrate the proposed methods, we consider a dataset from 1980 comprised of over 26,000 cancer-related deaths from over 47,000 demographic strata from the Commonwealth of Pennsylvania --- these data and the external information we consider ``publicly available'' are described in Section~\ref{sec:data}.   Synthetic data are then generated under various modeling assumptions and various privacy budgets in Section~\ref{sec:synthetic}.  Following a brief overview of the paper's key findings, we discuss the work's implications in the context of a ``Synthetic CDC WONDER'' and next steps in Section~\ref{sec:disc}.

\section{Methods}\label{sec:methods}
\subsection{Notation and definitions}
We let $y_i$ denote the number of events belonging to group $i$ out of a population of size $n_i$, for $i=1,\ldots, I$ and $I\ge 2$.  While each individual $y_i$ is deemed potentially sensitive, we assume $y_{\cdot}=\sum_i y_i>0$ is not sensitive and thus is publicly available; e.g., annual reports released by the CDC include the number of deaths due to major causes of death at the \emph{state} level (e.g., deaths due to cancer), but not at the county level \citep[e.g., Table~12 of][]{deaths:2017}. Furthermore, while the presentation used here indexes the data by a single subscript, in many settings (including that used in Section~\ref{sec:analysis}) it may be more natural to include \emph{multiple} subscripts to denote multiple subgroups (e.g., age, race, and sex).

Before describing the framework of \citet{quick:diffpriv} and our approach for prior predictive truncation, we begin by describing the general framework we use to draw synthetic data, $\bz$, from the posterior predictive distribution that satisfy $\eps$-differential privacy \citep{dwork:etal}.  First, we specify a distribution for the true data, $\by=\left(y_1,\ldots,y_I\right)^T$, given a collection of model parameters, $\bphi$, denoted $p\left(\by\given \bphi\right)$.  We then specify a prior distribution for $\bphi$ given a set of known hyperparameters, $\bpsi$, denoted $p\left(\bphi\given \bpsi\right)$, and obtain the posterior distribution $p\left(\bphi\given \by,\bpsi\right)$. From this, we can then sample $\bz$ from the posterior predictive distribution, $p\left(\bz\given \by,\bpsi\right) = \int p\left(\bz\given\bphi\right) p\left(\bphi\given \by,\bpsi\right) d\bphi$.  The data synthesis mechanism $p\left(\bz\given \by,\bpsi\right)$ is said to be $\eps$-differentially private if for all possible $\by$ and $\bz$ and for any hypothetical dataset $\bx=\left(x_1,\ldots,x_I\right)^T$ with $\Vert\bx-\by\Vert_1=2$ and $\sum_i x_i=\sum_i y_i$ --- i.e., where there exists $i$ and $i'$ such that $x_i=y_i-1$ and $x_{i'}=y_{i'}+1$ with all other values equal --- then
\begin{align}
\left\vert \log \frac{p\left(\bz\given \by,\bpsi\right)}{p\left(\bz\given \bx,\bpsi\right)}\right\vert \le \epsilon.\label{eq:dp_dfn}
\end{align}

The U.S.\ Census Bureau's OnTheMap tool was the first synthetic data production system based on differential privacy.  OnTheMap is a tool that provides access to synthetic data on commuting patterns of U.S.\ workers --- e.g., the number of residents of geographic region $A$ who commute to work in geographic region $B$, denoted $y_{A;B}$.  These synthetic data are generated using a multinomial-Dirichlet mechanism proposed by \citet{onthemap} which, under certain conditions, can satisfy~\eqref{eq:dp_dfn}.  Specifically, the approach assumes that the number of workers commuting to region $B$ from each of the various regions in the spatial domain, denoted $\by_B = \left(y_{1;B},\ldots,y_{I;B}\right)^T$, follows a multinomial distribution with $\sum_{i} y_{i;B} = y_{B\cdot}$ total events and probabilities denoted by $\theta_{i;B}$, and where $\btheta_{B} \sim \Dir\left(\balpha_B\right)$; synthetic data, $\bz_{B}$, are then sampled from the resulting posterior predictive distribution.  For $\bz_{B}$ to satisfy differential privacy for a given privacy budget, $\eps >0$, the hyperparameters $\alpha_{i;B}$ must be sufficiently large.  When $\eps$ is small, however, the requirements for the $\alpha_{i;B}$ become prohibitively high, resulting in a prior distribution which would dominate the data and thereby hinder the utility of the synthetic data \citep{charest:2010}.  The approach of \citet{onthemap} can be viewed as a special case of the approach of \citet{quick:diffpriv}, and thus a more detailed derivation of its properties will be discussed in the following subsection.

\subsection{The Poisson-gamma mechanism}\label{sec:pg}
When generating synthetic data the context of CDC WONDER --- e.g., the number of deaths due to a given cause of death in a given county --- there are two important factors that ought to be taken into account.  First and foremost, U.S.\ counties vary wildly in their population sizes, even within the same state --- e.g., the most populated county in the state of Texas (Harris County) is home to more than 4.7 million residents while the three least populated counties each have fewer than 500 residents.  In addition, death rates can vary substantially by cause of death and by demographic factors like age, race/ethnicity, and sex.  Thus, a synthetic data mechanism with the flexibility to account for heterogeneity in population sizes and/or event rates should be expected to out perform an otherwise comparable mechanism that fails to do so.  From this point forward, we assume information regarding group-specific population sizes, $n_i$, and suitable estimates of the underlying event rates are publicly available --- support for this assumption will be provided as part of the illustrative example in Section~\ref{sec:analysis}.

With this setup in mind, we follow the approach of \citet{quick:diffpriv} and let
\begin{align}
y_i\given\lambda_i\sim \Pois\left(n_i\lambda_i\right) \;\;\text{and}\;\;\lambda_i\sim \Gam\left(a_i,b_i\right),\label{eq:pglik}
\end{align}
where $\lambda_i$ denotes the underlying event rate in group $i$ and $a_i$ and $b_i$ denote group-specific hyperparameters such that $E\left[\lambda_i\given a_i,b_i\right]=a_i\slash b_i = \lambda_{i0}$ is known (i.e., based on publicly available information).  It is then straightforward to show that $\lambda_i\given y_i \sim \Gam\left(y_i+a_i,n_i+b_i\right)$.  In addition, since the $\lambda_i$ are conditionally independent, any weighted average of the $\lambda_i$ with unnormalized weights $n_i+b_i$ will also be a gamma random variable --- e.g., the weighted average of event rates \emph{not} associated with group $i$, denoted $\lambda_{(i)}$, can be written as
\begin{align}
\lambda_{(i)}\given \by_{(i)} \sim \Gam\left(y_{(i)}+a_{(i)}, n_{(i)}+b_{(i)}\right), \;\text{where}\; \lambda_{(i)} = \frac{\sum_{j\ne i} \left(n_j+b_j\right) \lambda_j}{\sum_{j\ne i} \left(n_j+b_j\right)}
\label{eq:noti}
\end{align}
and $y_{(i)}=\sum_{j\ne i}y_j$, with similar definitions for $a_{(i)}$, $n_{(i)}$, and $b_{(i)}$.  From~\eqref{eq:pglik} and~\eqref{eq:noti}, it can then be shown that the posterior predictive distributions for the synthetic counts, $z_i$ and $z_{(i)}$, will each take the form of a negative binomial distribution --- e.g., $z_i\given y_i,a_i,b_i \sim \NegBin\left(y_i+a_i,n_i\slash \left(b_i+2*n_i\right)\right)$.  Thus, if we wish to generate a vector of synthetic counts, $\bz_i = \left(z_i,z_{(i)}\right)^T$ such that $z_i+z_{(i)}=y_{\cdot}$, our interest lies the joint distribution, $p\left(\bz_i\given \by,\ba,\bb\right)$, which can be expressed as
\begin{align}
p\left(\bz_i\given \by,\ba,\bb\right)
&=\frac{
\frac{\Gamma\left(z_i+y_i+a_i\right)}{z_i!} \left(\frac{n_i}{b_i+2*n_i}\right)^{z_i} \times
\frac{\Gamma\left(z_{(i)}+y_{(i)}+a_{(i)}\right)}{z_{(i)}!} \left(\frac{n_{(i)}}{b_{(i)}+2*n_{(i)}}\right)^{z_{(i)}}
}{\sum_{z=0}^{y_{\cdot}} \frac{\Gamma\left(z+y_i+a_i\right)}{z!} \left(\frac{n_i}{b_i+2*n_i}\right)^{z} \times
\frac{\Gamma\left(y_{\cdot}-z+y_{(i)}+a_{(i)}\right)}{\left(y_{\cdot}-z\right)!} \left(\frac{n_{(i)}}{b_{(i)}+2*n_{(i)}}\right)^{\left(y_{\cdot}-z\right)} }. \label{eq:poiscond}
\end{align}
Thus, demonstrating that $p\left(\bz_i\given \by,\ba,\bb\right)$ satisfies $\eps$-differential privacy requires that we can establish bounds for the ratio
\begin{align}
\frac{p\left(\bz_i\given\by,\ba,\bb\right)}{p\left(\bz_i\given \bx,\ba,\bb\right)} =& \frac{C\left(\bx,\bn,\ba,\bb\right)}{C\left(\by,\bn,\ba,\bb\right)}
\times \frac{\Gamma\left(z_i+y_i+a_i\right)}{\Gamma\left(z_i+x_i+a_i\right)} \times
\frac{\Gamma\left(z_{(i)}+y_{(i)}+a_{(i)}\right)}{\Gamma\left(z_{(i)}+x_{(i)}+a_{(i)}\right)},\label{eq:pgdp1}
\end{align}
where $\bx_i=\left(x_i,x_{(i)}\right)^T$ represents a hypothetical dataset such that $\left\Vert\bx_i-\by_i\right\Vert_1=2$ and where
\begin{align}
C\left(\by,\bn,\ba,\bb\right) = \sum_{z=0}^{y_{\cdot}} \frac{\Gamma\left(z+y_i+a_i\right)}{z!}\frac{\Gamma\left(y_{\cdot}-z+y_{(i)}+a_{(i)}\right)}{\left(y_{\cdot}-z\right)!} \times r_i\left(\bn,\bb\right)^{z},\label{eq:pg_norm}
\end{align}
and $r_i\left(\bn,\bb\right)=\left(b_{(i)}\slash n_{(i)}+2\right)\slash \left(b_i\slash n_i+2\right)$.  As proven in \citet{quick:diffpriv} --- and as demonstrated in Appendix~A.1 --- in order for~\eqref{eq:poiscond} to satisfy $\eps$-differential privacy, we require
\begin{align}
a_i \ge \frac{y_{\cdot}}{e^{\eps}\slash \nu_i -1}, \; \text{where}\; \nu_i = \frac{y_{\cdot}\times\lceil 1-r_i\left(\bn,\bb\right)\rceil^{+}+a_{(i)}+y_{\cdot}-1}{a_{(i)}+y_{\cdot}-1}.
\label{eq:pgdp_a}
\end{align}
Furthermore, if~\eqref{eq:pgdp_a} holds for all $i$ simultaneously, then the mechanism imposed by~\eqref{eq:pglik} for $\bz=\left(z_1,\ldots,z_I\right)$ under the constraint that $\sum_i z_i = y_{\cdot}$, denoted $p\left(\bz\given\by,\ba,\bb\right)$, will satisfy $\eps$-differential privacy.  To sample from $p\left(\bz\given\by,\ba,\bb\right)$, we can leverage the fact that because the $y_i$ are conditionally independent given $\blambda$, we can write
\begin{align}
\by \given y_{\cdot},\blambda \sim \Mult\left(y_{\cdot},\bpi\right), \;\text{where}\; \pi_i=\frac{n_i\lambda_i}{\sum_{j=1}^I n_j\lambda_j}.\label{eq:pg_md}
\end{align}
Then if $\lambda_i^*$ is drawn from the posterior distribution, $\lambda_i\given y_i \sim \Gam\left(y_i+a_i,n_i+b_i\right)$, and if $\pi_i^*$ is defined as in~\eqref{eq:pg_md} as a function of $\lambda_i^*$, then sampling $\bz \given \bpi^*,y_{\cdot} \sim \Mult\left(y_{\cdot},\bpi^*\right)$ will be equivalent to a draw from $p\left(\bz\given\by,\ba,\bb\right)$.

If $n_i=n$ and $E\left[\lambda_i\given a_i,b_i\right]=\lambda_0$ for all $i$ --- i.e., if we assume all groups have the same population size and the same prior expected event rate --- then the Poisson-gamma mechanism of \citet{quick:diffpriv} is \emph{mathematically equivalent} to the multinomial-Dirichlet mechanism of \citet{onthemap} with $\balpha=\ba$.  While the Poisson-gamma mechanism addresses two key limitations of the multinomial-Dirichlet framework of \citet{onthemap} --- namely that it allows for heterogeneity in population sizes via the $n_i$ parameters and heterogeneity in the prior event rates via the hyperparameters, $a_i$ and $b_i$ --- one limitation both approaches share is that when the total number of events, $y_{\cdot}$, is large, the restriction in~\eqref{eq:pgdp_a} quickly becomes large for even moderate values of $\eps$.  In contrast, the methods proposed in Section~\ref{sec:trunc} aim to construct a framework in which the informativeness of the prior is primarily a function of $E\left[y_i\given \ba,\bb\right]$.

\subsection{Prior predictive truncation}\label{sec:trunc}
The reason the Poisson-gamma framework proposed by \citet{quick:diffpriv} suffers from low utility for moderate values of $\eps$ is that it is designed to protect against an extremely improbable worst case scenario --- i.e., where \emph{all} of the events in the synthetic dataset are assigned to a group in which only one event occurred.  To address this issue in the multinomial-Dirichlet setting, \citet{onthemap} proposed a framework referred to as $\left(\eps,\delta\right)$-probabilistic differential privacy in which the synthetic data would satisfy $\eps$-differential privacy with probability $\delta \ge 0$.  Aside from not satisfying \emph{pure} ($\delta=0$) differential privacy, a drawback of this approach is that because the hyperparameters, $\balpha$, are based on the \emph{posterior} predictive distribution, it requires an iterative approach to determine the optimal values of $\balpha$, and thus releasing these values may leak additional information about the data.

Here, we propose the use of \emph{prior} predictive truncation, specifically the use of $n_i$ and $\lambda_{i0}=E\left[\lambda_i\given a_i,b_i\right]$ to specify \emph{a priori} upper and lower bounds on $z_i$ taking the form:
\begin{align}
L_i = F^{-1}\left(\alpha\slash 2\given c^{-1}n_i \lambda_{i0}\right) \le z_i \le F^{-1}\left(1-\alpha\slash 2\given cn_i\lambda_{i0}\right) = U_i,\label{eq:zbounds}
\end{align}
where $F^{-1}\left(\cdot\given \mu\right)$ denotes the inverse cdf of a Poisson random variable with mean $\mu$ and where $\alpha \in \left(0,1\slash2\right)$ and $c\ge1$ represent tuning parameters to construct the bounds.  The objective here is \emph{not} to force $z_i \approx y_i$, but simply to restrict $z_i$ to a range of plausible values. Moreover, it should be emphasized that while it would be \emph{desirable} for $y_i \in \left[L_i,U_i\right]$, this is not a \emph{requirement} as that alone may leak information about $\by$ and thus be a violation of differential privacy.  That said, we \emph{will} truncate the posterior contribution of $y_i$ to the range $\left[L_i,U_i\right]$ --- e.g., any $y_i > U_i$ will be replaced by $U_i$ in~\eqref{eq:pglik}.  As a result, this approach is highly sensitive to the quality of the prior information --- e.g., if $y_i \gg E\left[y_i\given \ba,\bb\right] = n_i \lambda_{i0}$, then the proposed bounds may be too restrictive and thus may reduce the utility of the synthetic data.  A discussion of how tuning parameters like $\alpha$ and $c$ can be chosen is provided in Section~\ref{sec:disc}.

To implement the proposed prior predictive truncation, we begin by assuming that $E\left[y_i\given \ba,\bb\right] < E\left[y_{(i)}\given \ba,\bb\right]$ for all $i$ --- i.e., that no single group ``dominates'' the remaining groups.  While this assumption can certainly be violated in practice --- e.g., when a majority of a state's population belongs to a single urban center --- this assumption will tend to hold in scenarios in which the data are stratified by multiple factors (e.g., by spatial region \emph{and} demographic categories like age, race, and sex).  Nevertheless, Appendix~A.2 outlines how one can proceed when this assumption is violated.  Next, in order to demonstrate that this approach can be proven to satisfy $\eps$-differential privacy, we first restrict our focus to the ``group $i$ versus \emph{not} group $i$'' scenario, which --- as demonstrated in~(A.8) of Appendix~A.2 --- yields a bivariate distribution for $\bz_i=\left(z_i,z_{(i)}\right)^T$ of the form:
\begin{align}
\frac{p\left(\bz_i\given \by,\ba,\bb,{\cal B}\right)}{p\left(\bz_i\given \bx,\ba,\bb,{\cal B}\right)} =& \frac{C_{L_i}^{U_i}\left(\bx,\bn,\ba,\bb\right)}{C_{L_i}^{U_i}\left(\by,\bn,\ba,\bb\right)}
\times \frac{\Gamma\left(z_i+y_i+a_i\right)}{\Gamma\left(z_i+x_i+a_i\right)} \times \frac{\Gamma\left(z_{(i)}+y_{(i)}+a_{(i)}\right)}{\Gamma\left(z_{(i)}+x_{(i)}+a_{(i)}\right)}, \label{eq:truncrat}
\end{align}
where ${\cal B}=\left\{\left(L_i,U_i\right): i=1,\ldots,I\right\}$ denotes the set of bounds based on~\eqref{eq:zbounds} and
\begin{align}
C_{L_i}^{U_i}\left(\by,\bn,\ba,\bb\right) = \sum_{z=L_i}^{U_i} \frac{\Gamma\left(z+y_i+a_i\right)}{z!} \times
\frac{\Gamma\left(y_{\cdot}-z+y_{(i)}+a_{(i)}\right)}{\left(y_{\cdot}-z\right)!}  \times r_i\left(\bn,\bb\right)^{z}.\label{eq:ctrunc}
\end{align}
The key result of this paper is then summarized in Theorem~\ref{thm:crat} below:
\begin{thm}\label{thm:crat}
Let $C_{L_i}^{U_i}\left(\by,\bn,\ba,\bb\right)$ be as defined in~\eqref{eq:ctrunc} and $E\left[y_i\given a_i,b_i\right] \le E\left[y_{(i)}\given a_{(i)},b_{(i)}\right]$.  Then if $\bx=\left(x_i,x_{(i)}\right)^T$ and $\by=\left(y_i,y_{(i)}\right)^T$ denote vectors of non-negative integers of length 2 such that $x_i=y_i-1$ and $x_{(i)}=y_{(i)}+1$
and $x_i+x_{(i)}=y_i+y_{(i)}=y_{\cdot}$, then
\begin{align}
\frac{C_{L_i}^{U_i}\left(\bx,\bn,\ba,\bb\right)}{C_{L_i}^{U_i}\left(\by,\bn,\ba,\bb\right)} < \frac{y_{\cdot}-L_i+a_{(i)}+y_{(i)}}{L_i+a_i+y_i-1}. \label{eq:truncbound}
\end{align}
\end{thm}
\noindent
Thus, the expression in~\eqref{eq:truncrat} can be maximized by incorporating the bound from Theorem~\ref{thm:crat} and letting $y_i=L_i+1$ and $z_i=U_i$ to yield
\begin{align}
\frac{p\left(\bz_i\given \by,\ba,\bb,{\cal B}\right)}{p\left(\bz_i\given \bx,\ba,\bb,{\cal B}\right)} \le \frac{y_{\cdot}-L_i+a_{(i)}+y_{\cdot}-L_i-1}{L_i+a_i+L_i} \times \frac{U_i + a_i + L_i}{y_{\cdot}-U_i+a_{(i)}+y_{\cdot}-L_i-1}.\label{eq:truncalmost}
\end{align}
Thus, for $p\left(\bz_i\given \by,\ba,\bb,{\cal B}\right)$ to satisfy $\eps$-differential privacy, we require
\begin{align}
a_i \ge \frac{U_i - L_i}{e^{\eps} \slash \nu_i - 1} - 2*L_i, \;\text{where}\; \nu_i = \frac{y_{\cdot}-L_i+a_{(i)}+y_{\cdot}-L_i-1}{y_{\cdot}-U_i+a_{(i)}+y_{\cdot}-L_i-1}. \label{eq:pptdp}
\end{align}
Full details pertaining to the derivation of~\eqref{eq:pptdp} --- along with a proof of Theorem~\ref{thm:crat} --- are provided in Appendix~A.2.

While the requirement in~\eqref{eq:pptdp} is based on a bivariate set of synthetic data, $\bz_i$, requiring that $a_i$ satisfy the appropriate criteria from~\eqref{eq:pptdp} for all $i$ simultaneously will ensure that the mechanism for generating the complete vector of synthetic data, denoted $p\left(\bz\given \by,\ba,\bb,{\cal B}\right)$, will satisfy $\eps$-differential privacy.  The benefit of this approach is that the various $a_i$ are allowed to vary as a function of $E\left[y_i\given a_i,b_i\right]$, and thus groups with smaller expected event counts should be expected to receive \emph{far} less informative priors under~\eqref{eq:pptdp} than under~\eqref{eq:pgdp_a}, especially as the total number of events, $y_{\cdot}$, increases.  It should also be noted that because the expression for $a_i$ in~\eqref{eq:pptdp} is a function of $a_{(i)}$, we require the use of an iterative algorithm to simultaneously calculate the $a_i$ and $b_i$ parameters subject to the model's constraints --- see Appendix~B for more details.

\begin{figure}[t]
    \begin{center}
        \subfigure[Privacy Protections]{\includegraphics[width=.45\textwidth]{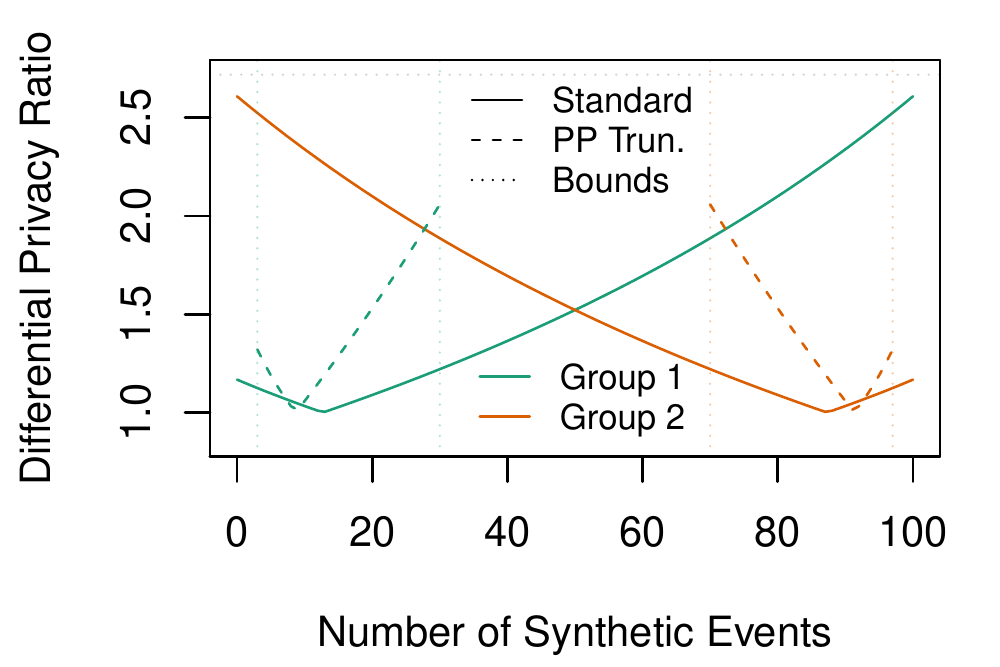}\label{fig:dp_fig}}
        \subfigure[Synthetic Data]{\includegraphics[width=.45\textwidth]{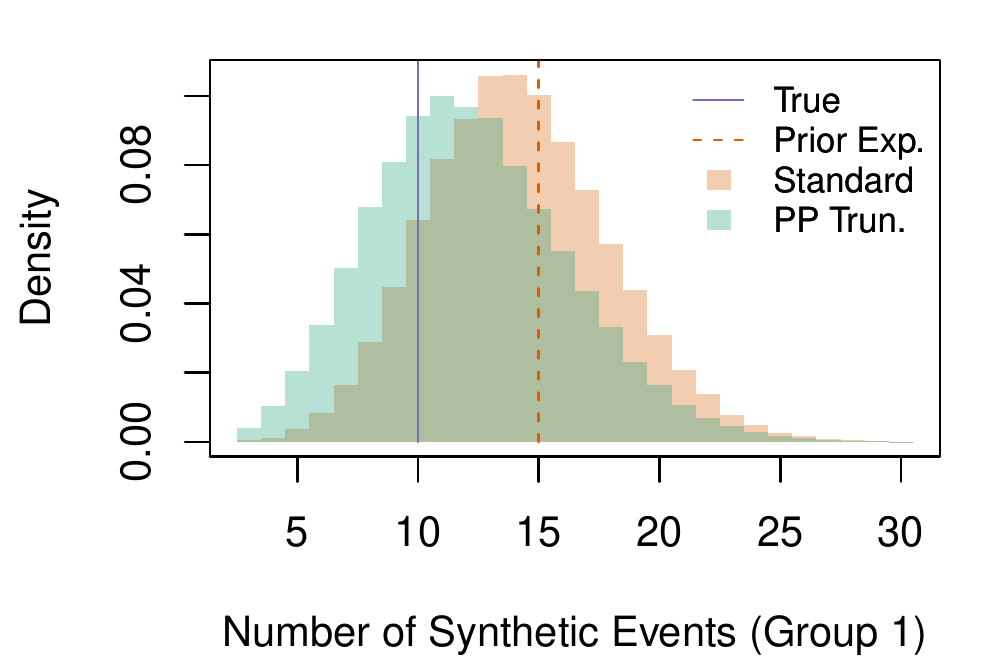}\label{fig:dp_syn}}
    \end{center}
    \caption{Generating synthetic data with $\eps=1$ for $y_{\cdot}=100$ events and $E\left[\by\given \ba,\bb\right]=\left(15,85\right)^T$.  In Panel~(a), the prior predictive truncation approach (dashed lines) yields less informative priors, which in turn cause the differential privacy ratios from~\eqref{eq:truncalmost} to increase at a much faster rate than under the standard framework (solid lines) based on~\eqref{eq:pgdp1} from \citet{quick:diffpriv}.  Panel~(b) compares the posterior predictive distribution for Group~1 from the competing approaches for sampling synthetic data --- for reference, the true value of $y_1=10$ and the prior expected value of $E\left[y_1\given \ba,\bb\right]=15$ are highlighted.}
    \label{fig:demo}
\end{figure}

As an illustration of how this approach works, we consider a dataset of $y_{\cdot}=100$ events belonging to $I=2$ groups with $E\left[\by\given \ba,\bb\right]=\left(15,85\right)^T$ with $\eps=1$.  As illustrated by the solid lines in Figure~\ref{fig:demo}, the approach of \citet{quick:diffpriv} maximizes the ratio from~\eqref{eq:pgdp1} when $\by=\left(1,99\right)^T$ and $\bz=\left(100,0\right)^T$.  While this maximum risk \emph{is} indeed less than the desired threshold of $\exp\left(\eps\right)=2.71$, the probability of allocating all $y_{\cdot}=100$ events to the group with $E\left[y_1\given\ba,\bb\right]=15$ is extremely low (i.e., less than $4.1 \times 10^{-83}$), and protecting against this worst case scenario requires $a_1>116$, a value which certainly will overwhelm whatever the true $y_1$ is.  In contrast, the proposed prior predictive truncation approach with $\alpha=10^{-4}$ and $c=1$ maximizes the ratio from~\eqref{eq:truncalmost} for Group~1 when $z_1=30$, as demonstrated by the dashed lines in Figure~\ref{fig:demo}.  It should be noted here that $z_1=30$ is twice as large as $E\left[y_1\given a_1,b_1\right]$, and thus is still a conservative bound.  Furthermore, the benefit of this approach is that we now only require $a_1>14.17$ to satisfy $\eps$-differential privacy.  Similarly, the requirement for $a_2$ is reduced from $a_2>58$ to just $a_2 > 0.001$.  As a result, we obtain posterior distributions for each $\lambda_i$ that are far more reliant on the true data than under \citet{quick:diffpriv} --- thereby producing synthetic data that are more similar to the true data and thus have greater utility --- without sacrificing data privacy.  Finally, it should be noted that the criteria specified in~\eqref{eq:pptdp} leaves some utility on the table in the sense that the upper bound from Theorem~\ref{thm:crat} is a \emph{conservative} upper bound, which results in the ratio from~\eqref{eq:truncalmost} in Figure~\ref{fig:dp_fig} being far less than $\exp\left(\eps\right)=2.71$ when $z_1=30$.  That said, while the requirements from~\eqref{eq:pptdp} may be suboptimal, they still produced an over eight-fold reduction in $a_1$ when $y_{\cdot}=100$, and --- because the requirements in~\eqref{eq:pgdp1} are primarily a function of $y_{\cdot}$ while the requirements in~\eqref{eq:pptdp} are a function of $U_i - L_i$ --- the benefits of this approach only grow as $y_{\cdot}$ increases.

\section{Illustrative Analysis}\label{sec:analysis}
\subsection{Pennsylvania cancer death data}\label{sec:data}
To illustrate the benefits of the prior predictive truncation described in Section~\ref{sec:methods} in a real-world application, we consider a dataset comprised of cancer-related death counts and population estimates from the Commonwealth of Pennsylvania (PA) in 1980.  We let $y_{icars}$ denote the number of deaths in county $i$ (of 67 counties) due to cancer type $c$ (of 9 types) from individuals belonging to age group $a$ (of 13 age groups, ranging from less than one year of age to 85 years and older), race $r$ (white, black, other), of sex $s$ (male/female).  Specifically, we consider cancers of the lip, oral cavity, and pharynx (ICD-9: 140--149), cancers of the digestive organs and peritoneum (ICD-9: 150--159), cancers of the respiratory and intrathoracic organs (ICD-9: 160--165), cancers of the breast (ICD-9: 174--175), cancers of the genital organs (ICD-9: 179--187), cancers of the urinary organs (ICD-9: 188--189), cancers of all other and unspecified sites (ICD-9: 170--173, 190--199), leukemia (ICD-9: 204--208), and all other cancers of the lymphatic and hematopoietic tissues (ICD-9: 200--203).  In total, there were $y_{\cdot}=\sum_{icars}y_{icars} = \text{26,116}$ cancer-related deaths in PA in 1980 belonging to these $67 \times 13 \times 9 \times 3 \times 2 = \text{47,034}$ strata, and because these data are from prior to 1989, they are publicly available via CDC WONDER free of suppression.  
Here, we assume that the total number of cancer-related deaths in PA, $y_{\cdot}$, is to be kept invariant (i.e., is safe to disseminate) but that the county-level death counts, $y_{icars}$, are to be considered sensitive.

The external information used in this analysis comes from two sources.  First and foremost, we incorporate the \emph{bridged-race population estimates} --- denoted $n_{iars}$ --- released annually by the National Center for Health Statistics.  As described by \citet{bridged:race}, these estimates are the product of a collaborative agreement between the CDC and the Census Bureau.  The bridged-race population estimates are model-based and are produced using data from the Census Bureau and the CDC's National Health Interview Survey.  These estimates were developed for the express purpose of creating denominators for vital rates by external researchers, thus we believe it is reasonable to assume that similar estimates will continue to be produced and publicly disseminated despite the Census Bureau's recent announcement \citep{census:dp} that differential privacy would be used to protect the 2018 End-to-End Census Test with an eye toward its use for the full 2020 Decennial Census of Population and Housing.  That said, it should be noted that it is entirely possible that future releases of the bridged-race population estimates could have differentially private protections built into their design which may result in a reduction in accuracy.

The second source of data used in this analysis comes from the CDC's annual National Vital Statistics Reports \citep[e.g.,][]{deaths:2017}.  Specifically, in addition to assuming that $y_{\cdot}$ is known, we assume that the cancer-related death rates \emph{at the national level} for each of the aforementioned strata are publicly known --- i.e., the national death rate for each combination of age, race, sex, and form of cancer, denoted $\lambda_{cars;0}$.  Due to deviations between overall rates at the state and national level, all strata-specific national death rates will be adjusted by a factor of $y_{\cdot}\slash \sum  n_{icars}\lambda_{cars;0}$.  To put these assumptions into perspective, analogous state-level totals for major causes of death (e.g., cancer) from 2017 are provided in Table~12 of the CDC's annual report, and national death rates by age and selected causes in 2017 are provided in Table~7 of the report and national death rates by race, sex, and selected causes in 2017 are provided in Table~8 \citep{deaths:2017}.  Previous iterations of the report (e.g., those prior to the inception of CDC WONDER) included death rates by age, race, sex, and selected causes (e.g., Tables~1-8 and~1-9 of NCHS, \citeyear{deaths:1980}) --- here, we operate under the assumption that the decision to discontinue publishing tables by age, race, sex, and selected causes was due to the inception of CDC WONDER (where such granular data is available) rather than due to privacy concerns.  Should the CDC determine that these national death reports must also be differentially private, we expect that the large denominators underlying these rates would provide stability to the sanitized values.

\begin{figure}[t]
    \begin{center}
        \subfigure[State vs.\ National Death Rates]{\includegraphics[width=.5833\textwidth]{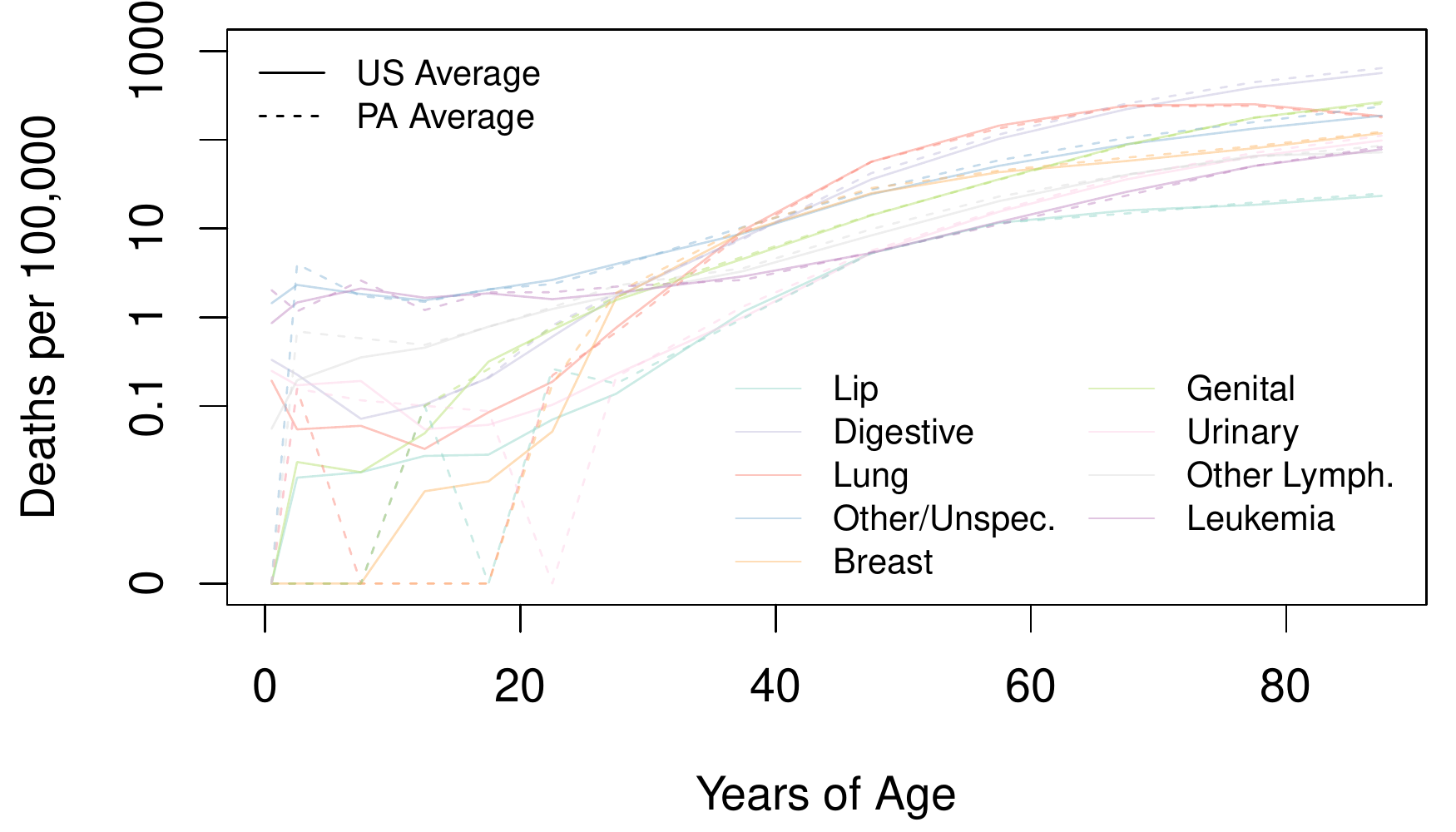}\label{fig:uspa_rates}}
        \subfigure[Observed vs. Expected Counts]{\includegraphics[width=.333\textwidth]{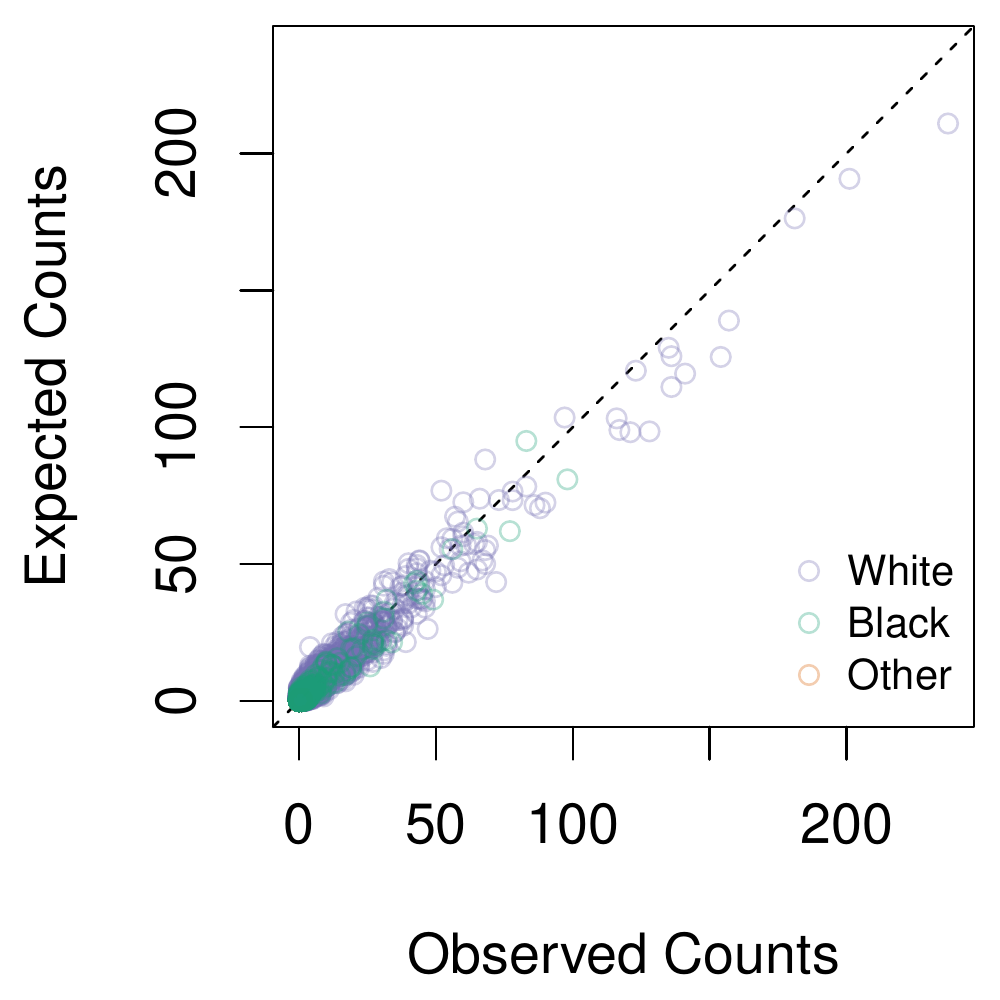}\label{fig:uspa_deaths}}
    \end{center}
    \caption{Cause-specific death rates at the national level and for the state of Pennsylvania.  National-level rates are used as prior information for estimating the proper allocation of deaths at the state and county level.}
    \label{fig:uspa}
\end{figure}

Figure~\ref{fig:uspa} demonstrates the utility of the prior information used in this analysis.  In Figure~\ref{fig:uspa_rates}, we compare the age/cause-specific death rates at the national level from 1980 and those at the state level, and in Figure~\ref{fig:uspa_deaths}, we compare the true death counts to the expected death counts based on the bridged-race population estimates and the national death rates.  In both cases, we see a high degree of agreement, as is desired.  That said, the decision to \emph{use} this information must be made \emph{prior} to utility checks of this form, as to do otherwise would result in a violation of differential privacy.  A discussion of how to choose sources of prior information and the implications on privacy protections is provided in Section~\ref{sec:disc}.

\subsection{Generating synthetic death counts}\label{sec:synthetic}
To generate differentially private synthetic data, we consider two approaches: the untruncated Poisson-gamma framework of \citet{quick:diffpriv} and the truncated Poisson-gamma framework proposed in Section~\ref{sec:trunc} with
$\alpha=1\slash\text{47,034}$ and $c=1$. For each approach, $L=\text{1,000}$ sets of synthetic data were generated for $\eps\in \left\{0.01,0.5,1,1.5,\ldots,4\right\}$.  Synthetic data generated from the untruncated approach of \citet{quick:diffpriv} were sampled using the approach outlined in Section~\ref{sec:pg}, and synthetic data generated from the proposed prior predictive truncation approach were sampled using the algorithm outlined in Appendix~C.

Before diving into a comparison of the synthetic data themselves, we first compare the prior distributions underlying each of the synthesizers.  When using the approach proposed by \citet{quick:diffpriv}, the prior distribution for all of our $y_{icars}$ required an informativeness of $a_{icars} > y_{\cdot}\slash \left(\exp\left(\eps\right)-1\right)$.  For instance, with $y_{\cdot}=\text{26,116}$ total deaths and $\eps=1$, this implies $a_{icars} > \text{15,000}$.  To put this in perspective, the largest value in our dataset is just $\max\left(y_{icars}\right)= 237$.  In contrast, using the prior predictive truncation approach described in Section~\ref{sec:trunc}, our priors all have $a_{icars} < 17$ with a median value of 0.58, values comparable to the untruncated approach with $\eps>7$.  As a result, we should expect the prior predictive truncation approach to put more weight on the observed data, thereby producing synthetic data with greater utility.  It should also be noted that (while not required) none of the $y_{icars}$ fell outside the bounds of the prior predictive truncation used in this analysis, suggesting that the prior information used is highly relevant and that the bounds constructed from our choice of $\alpha$ and $c$ were sufficiently conservative.

\begin{figure}[t]
    \begin{center}
        \subfigure[Original/Untruncated]{\includegraphics[width=.3\textwidth]{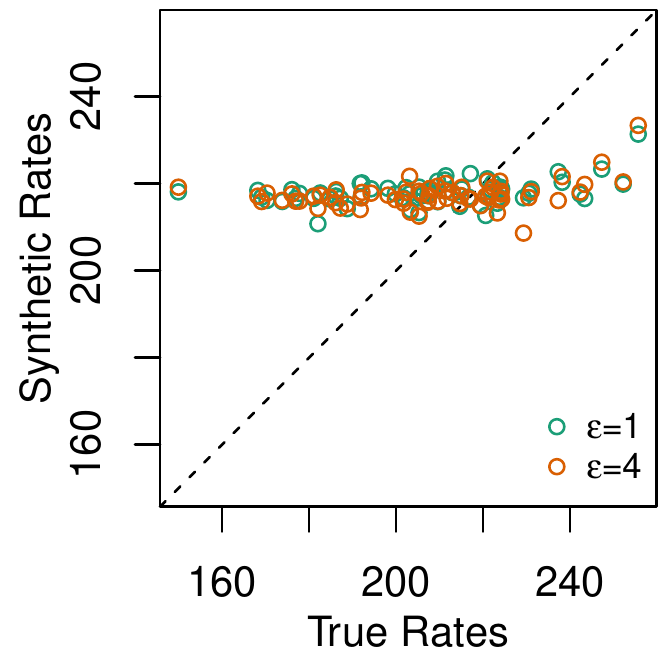}\label{fig:aa_old}}
        \subfigure[Prior Predictive Truncation]{\includegraphics[width=.3\textwidth]{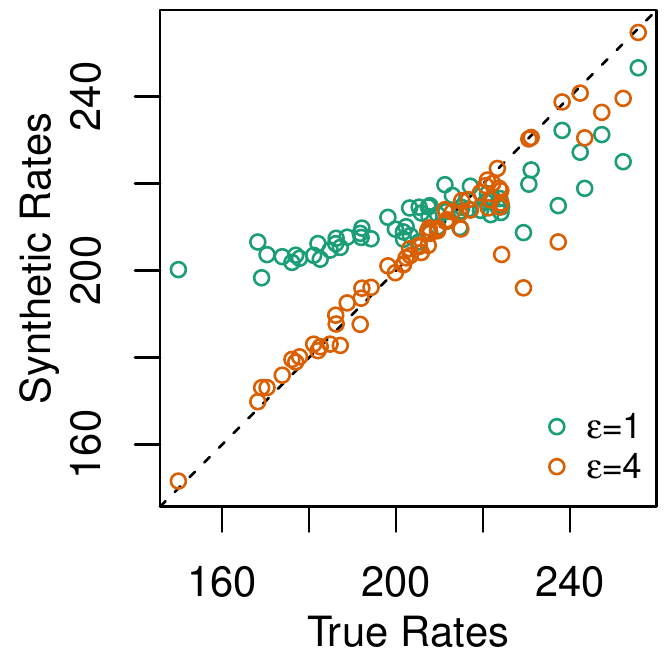}\label{fig:aa_trunc}}
    \end{center}
    \caption{Comparison of age-adjusted cancer-related death rates based on the two approaches for generating synthetic data for $\eps=1$ and $\eps=4$.}
    \label{fig:aa}
\end{figure}

\begin{figure}[t]
    \begin{center}
        \subfigure[True Age-Adjusted Rates]{\includegraphics[width=.36\textwidth]{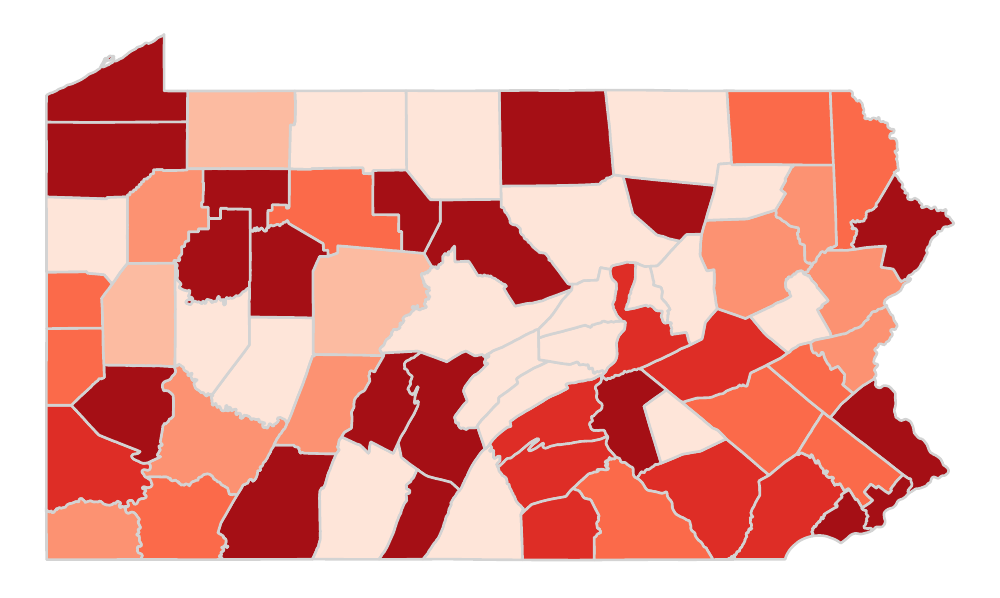}\label{fig:rate_true}}
        \subfigure[Synthetic Rates (PPT; $\eps=1$)]{\includegraphics[width=.36\textwidth]{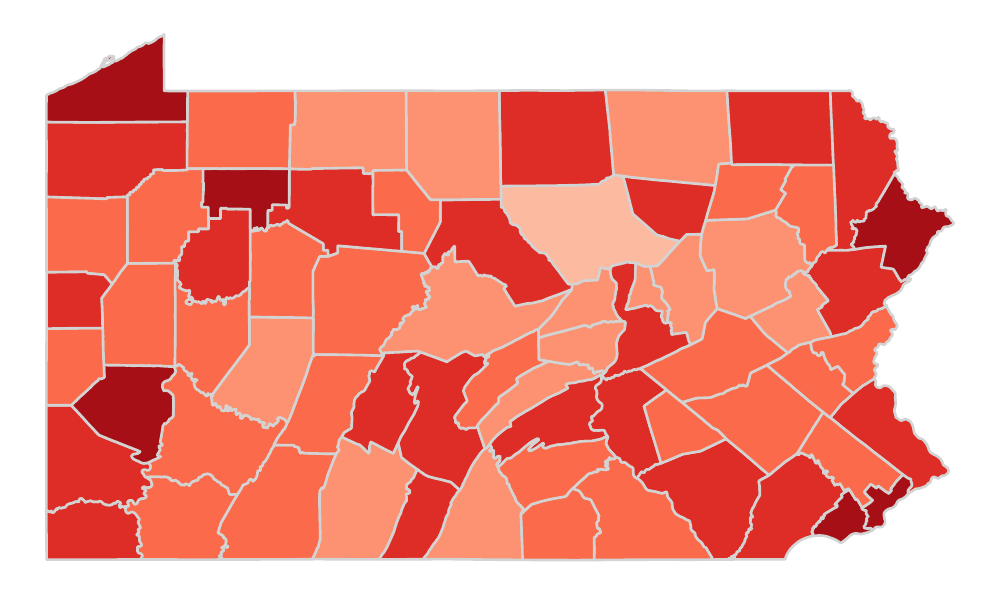}\label{fig:rate_trunc2}}
        \includegraphics[width=.12\textwidth]{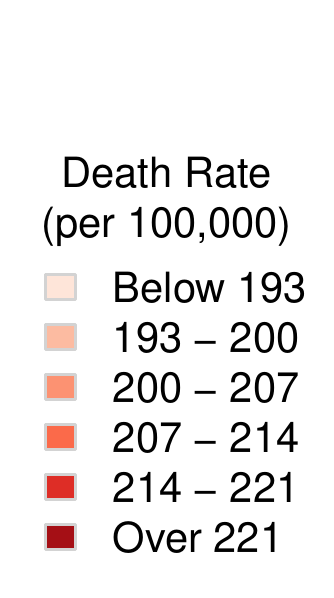}
    \end{center}
    \caption{Maps of the age-adjusted cancer death rates based on the true data and synthetic data generated under the proposed prior predictive truncation approach with $\eps=1$.}
    \label{fig:ratemaps}
\end{figure}

As an initial evaluation of the utility of the synthetic data, Figure~\ref{fig:aa} compares the age-adjusted death rates due to cancer (averaged over all $L$ sets of synthetic data) of both approaches for $\eps=1$.  Immediately, we see the effect of the highly informative priors imposed by the untruncated approach of \citet{quick:diffpriv}, as the rates in all $67$ counties have been pulled toward a common value of 217 deaths per 100,000 --- i.e., the national age-adjusted death rate.  In other words, the geographic variation in rates that are present in the true data has all but been smoothed away.  In contrast, the synthetic data based on the prior predictive truncation approach preserve geographic disparities in rates when $\eps=4$ and maintain \emph{some} heterogeneity in the rates when $\eps=1$.  This is more apparent in Figure~\ref{fig:ratemaps}, which compares the map of the true age-adjusted rates to the averages based on the synthetic data under the proposed prior predictive truncation approach when $\eps=1$.  Here, we see that while the synthetic data have more moderate rates throughout the less populated, rural counties, rates in PA's population centers are allowed to deviate from the norm.  Maps for other values of $\eps$ for the truncated model are provided in Figure~D.3 of the Web Appendix.  Due to the disparate levels of informativeness between the truncated and untruncated models (which can best be observed in Figure~D.1 of the Web Appendix), from this point forward we restrict our focus to the results from the proposed prior predictive truncation approach.

\begin{figure}[t]
    \begin{center}
        \subfigure[Urban/Rural Disparity]{\includegraphics[width=.45\textwidth]{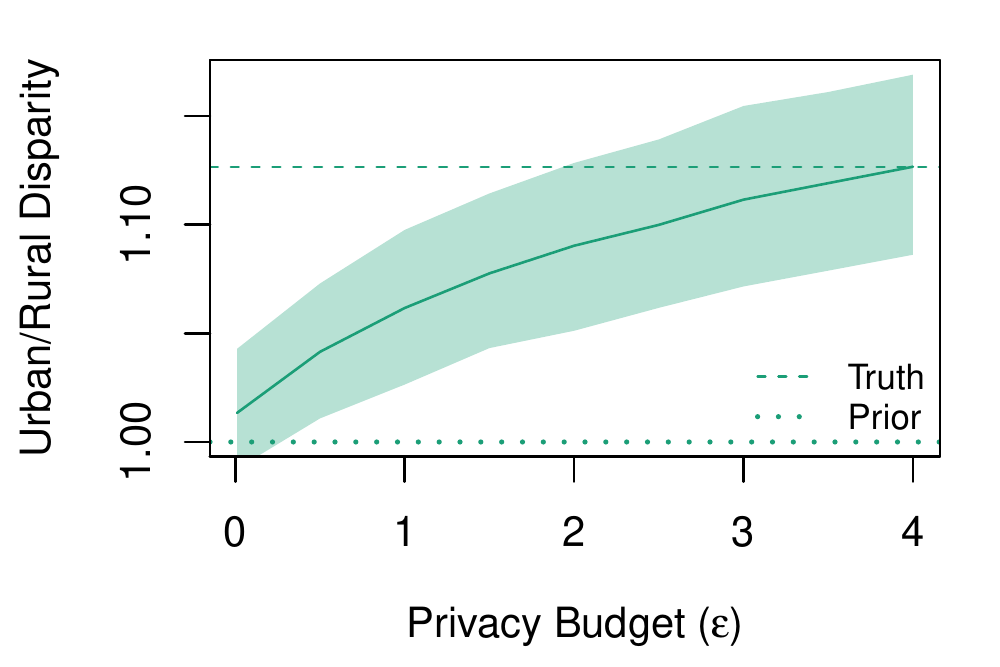}\label{fig:urb_rural}}
        \subfigure[Black/White Disparity]{\includegraphics[width=.45\textwidth]{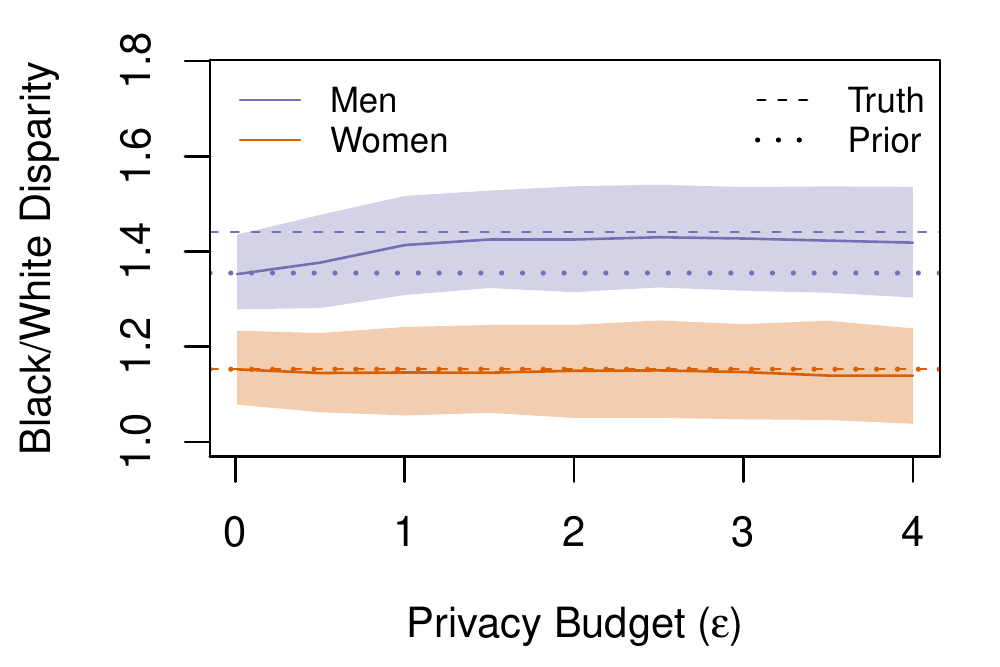}\label{fig:black_white}}
    \end{center}
    \caption{Estimated urban/rural disparities and black/white disparities (by sex) based on the synthetic data generated from the posterior predictive truncation approach for various levels of $\eps$.  Values based on the true data (dashed lines) and the prior information (dotted lines) are provided for reference.}
    \label{fig:disparities}
\end{figure}

We now shift our focus to two commonly investigated disparities in rates: disparities in rates between urban and rural counties and racial disparities.  For our assessment of urban/rural disparities, we begin by defining a county as being ``urban'' if its population density exceeded 280 persons per square mile and ``rural'' otherwise.  Based on this definition, 19 of PA's 67 counties are deemed ``urban'', and the age-adjusted cancer death rates in these counties were 15\% higher than those in rural counties.  In contrast, because the CDC's annual death reports \emph{do not} break down death rates by urban/rural status, our prior information assumes \emph{no disparities} exist between urban and rural counties.  The result of these two competing phenomena can be seen in the urban/rural disparity estimates produced by the synthetic data shown in Figure~\ref{fig:urb_rural}.  In particular, while the synthetic data yield accurate estimates of the urban/rural disparity when $\eps$ is large, the estimates gradually shift toward the null value from the prior as $\eps$ decreases and more informative priors are used.

Similar results are obtained in an investigation of the black/white disparities in cancer death rates, as shown in Figure~\ref{fig:black_white}.  In the true death data, cancer death rates for black men were nearly 45\% higher than for their white counterparts, while rates for black women were 15\% higher than for white women.  Unlike the urban/rural disparities example, however, our prior information based on the CDC's annual death reports \emph{do} contain information on racial disparities at the national level.  By coincidence, the black/white disparity in cancer death rates among women in the Commonwealth of Pennsylvania in 1980 was nearly identical to that at the national level, thus in this case our prior information effectively equals the true value.  It should be noted that this is \emph{not} to be expected, nor does it imply that the racial disparities at the \emph{county level} are also equivalent to those at the national level.  Thus, the takeaway message from Figure~\ref{fig:disparities} should not be that the prior predictive truncation approach will \emph{preserve} inference on quantities such as urban/rural disparities or racial disparities, but rather that the synthetic data should be expected to produce estimates that are bounded between the value from the true data (which will be unknown to data-users) and the publicly available national estimates, instead of yielding spurious associations.

While most of the results we have discussed thus far have been based on age-adjusted death rates for all types of cancer combined, it should be emphasized that the synthetic data in question were generated for each combination of age, race, sex, and specific type of cancer for each county.  To this end, additional results such as investigations of geographic, urban/rural, and racial disparities in death rates for specific forms of cancer are provided in Figures~D.4--D.7 of the Web Appendix.  Results for these subgroup analyses behave similar to the analyses presented in Figures~\ref{fig:aa}--\ref{fig:disparities} in that when $\eps$ is large, the synthetic data essentially provide the same inference (on average) as the true data, and as $\eps$ decreases, inference based on the synthetic data slowly begin to drift toward the values from the prior distribution.

\section{Discussion}\label{sec:disc}
\citet{quick:diffpriv} generalized the important work of \citet{onthemap} for generating differentially private synthetic event counts from a multinomial-Dirichlet framework to a Poisson-gamma model, and in doing so allowed users to incorporate public knowledge about heterogeneity in population sizes and underlying event rates to improve the utility of the synthetic data.  One limitation of the approach of \citet{quick:diffpriv} was that it was designed to protect against unrealistic worst-case scenarios --- e.g., the possibility of allocating all of our synthetic events to groups in which few events truly occurred.  To address this issue, we proposed the use of prior predictive truncation to restrict the range of values that the synthetic counts can take based on their prior predictive expected values.  In doing so, the gamma priors used in this framework are significantly less informative than those proposed by \citet{quick:diffpriv}, yielding a posterior distribution which leans more heavily on the true data and thus produces synthetic data with greater utility.

To address this issue in their own work, \citeauthor{onthemap}\ proposed an $\left(\eps,\delta\right)$-probabilistically differentially private approach in which the synthetic data generated from their model would satisfy $\eps$-differential privacy with probability $1-\delta$.  As with the approach proposed here, this allowed their approach to rely on less informative Dirichlet priors and thus improved the utility of the synthetic data.  While the proposed approach and the framework of \citet{onthemap} are similar, there is one key distinction between the two approaches: Because the approach of \citeauthor{onthemap}\ depends on the probability of sampling values of $\bz$ that violate $\eps$-differential privacy, the hyperparameters underlying their Dirichlet priors, $\balpha$, are a function of the true data and thus cannot be publicly disclosed without violating $\eps$-differential privacy.  In contrast, the restrictions imposed on $\bz$ described in Section~\ref{sec:trunc} are --- by definition --- based on the prior predictive distribution and thus are based solely on \emph{publicly available} information.  As a result, the restrictions imposed on $\bz$ and the values of the hyperparameters, $\ba$ and $\bb$, can be released without leaking sensitive information about the true data.  Previous work \citep[e.g.,][]{charest:2010} has considered treating synthetic data as noisy versions of the truth and using measurement error models (informed by the true hyperparameters) in an attempt to remove the differentially private noise and recover the true data --- while unexplored here, disclosing hyperparameters like $\ba$ and $\bb$ should help facilitate analyses of this nature.

While the proposed approach is capable of producing differentially private synthetic data with high utility, there is one key caveat --- the utility of the synthetic data under this approach \citep[and its predecessor, the approach of][]{quick:diffpriv} strongly depends on the quality of the prior information.  As a result, the use of the Poisson-gamma model with prior predictive truncation proposed here should only be considered in settings where high quality prior information is available and can be used with confidence.  For instance, \citet{quick:diffpriv} demonstrated that the the multinomial-Dirichlet framework of \citet{onthemap} can perform poorly in applications like CDC WONDER where a high degree of heterogeneity in population sizes and underlying event rates can occur, as these properties are in conflict with the framework's (implicit) assumption of homogeneity in the prior expected values.  Similarly, the prior information used in Section~\ref{sec:analysis} assumes that prior event rates are uniform across geographic areas --- while an assumption of this nature may suffice for synthesizing deaths due to \emph{chronic} diseases like heart disease and cancer, accounting for disparities between urban and rural regions may be crucial for other causes of death such as drug overdoses \citep{overdose_deaths} and infectious diseases \citep[e.g.,][]{covid:spatial}.  As such, the approach proposed here is \emph{not} intended for use as a black-box algorithm but rather an approach to be used in a very deliberate manner.  To this end, we believe agencies interested in using this approach should first identify historical data that can be used as a testbed to calibrate the framework for future data releases --- e.g., \emph{What sources of prior information are available?}, \emph{What ``known'' features are unaccounted for in the prior information?}, and \emph{What levels of the tuning parameters from~\eqref{eq:zbounds}, $\alpha$ and $c$, should be used?}  For instance, larger values of $c$ may be required when large disparities are expected but unaccounted for in the prior distributions (e.g., urban/rural disparities in death rates).  Choosing a value for $\alpha$ may be a bit more straightforward, with $\alpha< 1\slash I$ as a default rule-of-thumb.

Finally, we would be remiss to not acknowledge that conventional approaches for satisfying differential privacy are not based on posterior predictive synthesis as described here.  For instance, approaches such as the Laplace mechanism \citep{dwork:etal} and the geometric mechanism \citep{ghosh:geometric} operate by adding noise to the true counts. While the objective of this paper was to demonstrate the improvement in utility associated with prior predictive truncation, comparisons between the Poisson-gamma framework of \citet{quick:diffpriv} --- with and without truncation --- and existing, more conventional approaches for satisfying differential privacy must be conducted before it should be implemented in practice.  In particular, investigations of the performance of these various methods in producing accurate estimates of quantities explored here such as the geographic, urban/rural, and racial disparities in death rates --- this is an area of important and on-going research.

\bibliographystyle{jasa}
\bibliography{cdc_ref,cdc_epi,reports,disclosure,wonder}

\end{document}